**Dendritic organic electrochemical transistors grown by electropolymerization for 3D neuromorphic engineering**

*Kamila Janzakova, Mahdi Ghazal, Ankush Kumar, Yannick Coffinier, Sébastien Pecqueur\*, and Fabien Alibart\**

K. Janzakova, M. Ghazal, Dr. A. Kumar, Dr. Y. Coffinier, Dr. S. Pecqueur, Dr. F. Alibart
Institut d'Électronique, Microélectronique et Nanotechnologies (IEMN) – CNRS UMR 8520 – Université de Lille, boulevard Poincarré, 59652 Villeneuve d'Ascq, France.
E-mail: sebastien.pecqueur@univ-lille.fr; fabien.alibart@univ-lille.fr;

Dr. F. Alibart
Laboratoire Nanotechnologies Nanosystèmes (LN2) – CNRS UMI-3463 – 3IT, Sherbrooke J1K 0A5, Canada

One of the major limitation of standard top-down technologies used in today's neuromorphic engineering is their inability to map the 3D nature of biological brains. Here, we show how bipolar electropolymerization can be used to engineer 3D networks of PEDOT:PSS dendritic fibers. By controlling the growth conditions of the electropolymerized material, we investigate how dendritic fibers can reproduce structural plasticity by creating structures of controllable shape. We demonstrate gradual topologies evolution in a multi-electrode configuration. We conduct a detail electrical characterization of the PEDOT:PSS dendrites through DC and impedance spectroscopy measurements and we show how organic electrochemical transistors (OECT) can be realized with these structures. These measurements reveal that quasi-static and transient response of OECTs can be adjust by controlling dendrites' morphologies. The unique properties of organic dendrites are used to demonstrate short-term, long-term and structural plasticity, which are essential features required for future neuromorphic hardware development.

## 1. Introduction

Neuromorphic engineering is today attracting lots of attention as a potential direction to sustain artificial intelligence revolution.[1,2,3] Indeed, if standard machine learning approaches have reached unprecedented performances for complex computing tasks, their hardware implementation remains inefficient in term of energy consumption. More importantly, there is no clear strategy to date that could bring their energy performances at the level of biological systems for an equivalent computing performance. Alternatively, neuromorphic computing and engineering propose to explore further the different ingredients used by biological systems to compute and to reproduce them in hardware for unlocking this energy consumption limitation. [4,5] Neuromorphic approaches rely on spike-based encoding of information and bio-inspired learning rules for computing time dependent signals generated from different sensory modalities.[6] There is already great successes in this field with the recent neuromorphic chips[7] that have demonstrated record energy consumption, but there still exists several challenges that required the development of new solutions.[3,5,8] Firstly, biology relies on a large variety of information's carriers. Signals are transmitted and processed by the combination of several ions and various chemical messengers acting in concert. For instance, neuronal action potentials result from the ion gating that features various metallic cations ($K^+$, $Na^+$, $Ca^{++}$) and inorganic anions ($Cl^-$, $HCO_3^-$) mostly among many more ion traces information carriers.. Additionally, synaptic transmission is based on various neurotransmitters dynamics in between the pre- and post-neurons such as neurotransmitters binding rates and recovery, with additional contribution from heterosynaptic elements such as glial cells. [9] Such complex mechanisms provide a rich set of temporal dynamics for signal processing that are a key ingredient of biology for computing time dependent signals. [10] At the opposite, conventional neuromorphic hardware (i.e. CMOS-based technology) rely on electronic charges as a single carrier of information. This result in complex silicon circuits used to emulate each ingredient from biology individually.

Consequently, neuromorphic hardware needs to balance hardware complexity with biological realism, which can be a limitation for neuromorphic engineering. [7] Secondly, neural network architectures in biology are 3D. This gives to biology an incredibly large amount of parallelism for engineering neural circuits, which is evident from the rough estimate of one pre-neuron projecting its signal on 10000 post-neurons. Reaching this level of parallelism remains very challenging with 2D technologies that are quickly reaching interconnection's limitations.

Here, we capitalize on the recent progresses of organic mix iono-electronic conductors (OMIECs) for implementing innovative neuromorphic devices and functions. In contact with an electrolyte, this class of materials embeds both ionic transport through their ion-permeable bulk and electronic transport through their $\pi$-conjugated chains.[11] OMIECs have been the foundation for organic electrochemical transistors (OECTs) that use ionic gating of the electronic current through doping / de-doping of the organic conductor. Interestingly, ionic dynamics in OECTs as well as their intrinsic volumetric capacitance are providing rich dynamics in their temporal response for mimicking dynamics of synaptic plasticity. Along this line, OECTs have been used to demonstrate various synaptic plasticity mechanisms, from short-term memory, paired-pulse facilitation, long-term memory or dendritic integration.[12] More recently, electropolymerization of OMIECs has been used to implement long-term potentiation. [13] In this later case, the physical mechanism at work is oxido-reduction of monomers into polymers for creating the synaptic connections. While reversibility of the oxido-reduction remains challenging, some solutions exist for dissolving or over-oxidizing the electropolymerized material to ensure long-term depression. This approach has been realized in various 2D configurations, from thin-films, dendritic-like structures to wire-like. [14, 15]

In this paper, we show how complex iono-electronic dynamics in OMIECs as well as electropolymerization and oxido-reduction can be further exploited to implement a rich set of synaptic plasticities. We first evidence how 3D dendritic PEDOT:PSS fibers grown by bipolar electropolymerization can replicate structural plasticity observed in between neurons during the formation of complex networks. Secondly, we show that dendritic fibers can behave as OECTs and how their morphologies are influencing in depth their device behavior. We combine electrical DC characterization with electrochemical impedance spectroscopy to extract the key electrical parameters of dendritic OECTs. Additional 3D modeling of the dendritic structure is also used to explain the evolution of the OMIECs properties with electropolymerization conditions. Finally, these new material and devices properties are used to implement synaptic memory on various time scale, from permanent to short-term in electropolymerized dendritic network of PEDOT:PSS.

## 2. Results and discussion

### 2.1. Bipolar electropolymerization of structural plasticity implementation

Electropolymerization of PEDOT from EDOT monomers has been largely investigated as a bottom-up technique for material engineering.[17] This technique results in a large panel of material properties depending on the oxido-reduction reactions involved in the electropolymerization, electrolyte composition and electrical parameters use to drive the electropolymerization's reaction. More recently, bipolar electropolymerization of PEDOT was proposed for conductive fibers engineering.[14] These different works have shown the possibility to grow PEDOT in various configuration, from thin-film, dendritic-like to wire-like structures in between metallic electrodes on 2D substrates. We have recently shown how electrical parameters can be used to control dendritic structures grown in 3D in between two Au wires.[15] Here, bipolar electropolymerization setup consisted in two gold wires immersed in an aqueous solution of 10 mM EDOT monomers, 1 mM PSS:Na and 10 mM benzoquinone (figure 1a). Bipolar square-shape signal of 10 $V_{pp}$ and frequency $f_p$ was applied in between the two electrodes resulting during one half-cycle in oxidation of EDOT at one side and reduction of benzoquinone on the other side (opposite reaction occurring during the second half cycle). Figure 1b represents a typical time evolution of the dendrite growth during electropolymerization. With increase in time, dendrites grew from each electrode and touched each other to form multiple connections. We used frequency as a control parameter for tuning the shape of the dendritic PEDOT. Figure 1c shows the different dendritic structures obtained by increasing the electropolymerization frequency $f_p$ from 40 to 320 Hz. Larger frequency resulted in more wire-like structures in between the two electrodes while lower frequency favor more branches with larger sections. Based on our mesoscale modeling, a charge-particle driven mechanism suggests that this effect could be due to higher charge particle distribution at the center of both the electrodes. For a signal with short time period (high frequency), the charge particles from the center of the electrodes can only approach and attach to the tip of electrode, favoring wire like morphology. While for long time period (low frequency) the particles can transport over longer distances from the center of the electrodes, enabling fractal-like growth with multiple branches.[16] This effect can also be explained by the monomer permeation through the anodic double layer that forms partially depending on the applied frequency responsible for different dendrites' growth For high frequency and short time duration, the double layer can form only near the tip, while at low frequency, the double layer is capable to be formed isotropically, leading to high dendrite growth. [15] These different shapes resulted in well distinguishable conductance when a linear sweeping of voltage was applied in between the two electrodes. The linear current-voltage characteristics were a direct evidence of ohmic contact between the Au wires and the PEDOT:PSS fibers (figure 1d). Larger section and more numerous branches naturally resulted in larger conductances for lower $f_p$, which can be seen as a first level of analog programming of the conductance between the two terminals through dendritic shape engineering. Such creation of dendritic connections in between the two terminals (i.e. equivalent to the pre- and post-neurons) is reminiscent of the ability of biological neurons to establish connections with neighboring cells. During network formation, neural cells activity is known to favor preferential branching in between active cells.[18] Such structural plasticity mechanism presents the main advantage of limiting the memory resources for building network topologies where only useful connections are created. At the opposite, standard top-down approaches require to define a-priori all possible connections and to set non-useful synaptic weights to zero, which implies an over-estimate of the memory required for constructing artificial neural networks. Additionally, this structural plasticity is not limited to a binary connection and dendritic shape control can result in analog weight definition.

Figure 1e and 1f present a scanning microscope image of the dendritic PEDOT fibers grown at 80 Hz. The granular structure of the electropolymerized material with grain size in the micrometer range was in agreement with the proposed mechanism consisting in successive oxidation of monomers into oligomers before attaching to the surface. To evaluate the conductivity of the electropolymerized PEDOT material, we modeled the dendritic structure with a simple cylinder model reconstructing the 3D aspect of the 2D images (see Supplementary Informations figure S1 and figure S2). As the wire width varies across the dendrite morphology, the morphology is described as a collection of multiple cylinders, each having a diameter equal to the morphology's local width. Figure 1g shows the resulting current density map obtained from the 3D reconstructed dendrite images. The theoretical resistance of the resultant image was first computed. Then, resistivity for different dendrite morphologies based on the ratio of the experimental and the theoretical resistance values was calculated (figure 1g). Bipolar electropolymerization frequency resulted in a drop of resistivity for larger $f_p$ that could be associated to different organizations and composition of the PEDOT:PSS domains. Interestingly, material engineering with bipolar electropolymerization can result in different intrinsic electronic properties of the OMIEC material for device engineering.

## 2.2. Heterosynaptic plasticty of dendrictic PEDOT fibers

Such structural plasticity presents the drawback of being irreversible. Synaptic plasticity in biological networks is known to present both depression and potentiation during learning. Synaptic weight adjustment can be either controlled through homosynaptic mechanisms (i.e.

synaptic weight are only dependent on pre- and post-neuron activities) or heterosynaptic mechanisms (i.e. synaptic weight modulation in between pre- and post-neurons is controlled by an external element such as glial cells). This later effect has been recently evidence as an additional key element of learning in neural networks.[19] To mimic such heterosynaptic plasticity, we subsequently considered dendritic PEDOT connections as OECT devices. OECTs behavior is based on the doping / dedoping of the PEDOT material when ions are injected / extracted into / from the OMIEC material by an external gate potential. Here, the source (S) and drain (D) electrodes corresponded to the two Au wires connecting the dendritic fiber and the gate (G) electrode was an external Ag/AgCl electrode immersed into the electrolyte (figure 2a). Transfer characteristics were measured in between – 0.1V and 0.1 V in order to prevent any further electropolymerization reaction but to ensure ionic gating only of the PEDOT:PSS (EDOT oxidation reaction and BQ reduction reactions are expected to occur at potentials of 0.816 V and -0.105 V). Dendritic fibers presented effective OECT behavior with both depletion mode in the positive polarity and accumulation mode in the negative polarity (figure 2b).[20] This behavior could be associated to partial PEDOT doping with PSS-.We extracted the transconductance (see Supplementary Information, figure S3) from the transfer characteristics at 0V. The non-monotonic trends of the transconductance with $f_p$ can be understood at the light of the previous electrical characterizations and simulations. Transconductance in OECT is proportional to μ.C with μ the effective electronic mobility and C the total electrical capacitance of the OMIEC material (i.e. C being proportional to the amount of ions to be injected / repealed at a given gate potential). In dendritic fibers, it is straightforward to expect an increase of the bulk capacitance C at lower $f_p$ since the absolute amount of material is larger (note that volumetric capacitance is expected to hold for electropolymerized PEDOT). This would imply a monotonic increase of transconductance when decreasing $f_p$. In the other hand, 3D electrical modeling has evidenced a decrease of conductivity when decreasing $f_p$. This translates into a decrease of mobility μ when $f_p$ is decreased, which induces a decrease of transconductance.

Both C and μ were contributing in opposite direction to transconductance and resulted in the non-monotonic evolution presented in figure 2c. In other words, transconductance was limited by effective mobility at low $f_p$ and by capacitance at higher $f_p$. Overall, external gate control can be considered as an interesting opportunity to tune the dendritic connections' conductivity, as heterosynaptic mechanisms regulate synaptic conductances in biological networks.

In addition to static transfer characteristics, we performed electrochemical impedance spectroscopy to gain insight on the capacitive properties of the dendritic OECTs. Figure 2d reports the impedance modulus evolution with frequency in the low signal regime ($V_{apply} = 20$ mV) for both conventionnal OECTs realized by spin-coating of PEDOT:PSS and dendritic OECTs. Both devices were fitted with the electrical model shown in figure 2d. With constant phase element (CPE) that models the electrochemical charge/discharge behavior of the ionic device, $R_a$ the resistance of the device, 1/G the resistance of the electrolyte (which depends on the electrolyte's conductivity) and C parallel to $R_b$ representing the capacitance and the resistance of the Ag/AgCl gate electrode. Notably CPE index close to unity was the signature of quasi-ideal RC capacitive element from the conventional OECTs and CPE index $0,51 < n < 0,55$ was the signature of non-ideal capacitive response for dendritic OECTs.

Electrochemical impedance models provide a broader definition of an electrode charge/discharge kinetics for transient currents that do not empirically decay as an exponential of time upon step-voltage electrode polarization. Because ionic charge carriers have a non-negligible mass (at the opposite of electrons in a semiconductor) and also because redox processes generate or consume matter at the vicinity of the charged electrodes, diffusion and Faradically-driven mass transfers account additively to electrostatic forces upon operation of a device governed by such elementary mechanisms. This yields to unique electrochemical impedimetric signatures such as the one of CPE.[21] CPE are defined as passive dipole elements for which the spectral definition of their admittance is generically defined as $1/Z = Q_0*(i\omega)^n$. Their analytical expression showing a power of i ($i^2=-1$) with n other than 0 or $\pm1$ implies that their expression in the time domain can add new functionalities than RLC elements' which depend only on two factors: $Q_0 > 0$ and $0 < n < 1$. Moreover, they relate to fractional derivatives of time for which the contribution has been shown to be an essential ingredient in some neural models to implement longer-term volatility in the memory with power law dynamics.[22] Such phenomenon can arise from non-ideal charging/discharging of an electrode/electrolyte interface that generically observes a non-uniform electric field distribution across an electrode or all-along it.[23] In the case of standard PEDOT:PSS OECT showing both a potential distribution along the source-drain channel and accross the gated-material thickness, a minor contribution from the Warburg element (CPE with $n = 0.5$) can be observed in the Niquist plot, in series with the main bulk capacitance.[24] At the opposite, the main contribution in impedance spectra of dendritic OECT was associated to the CPE element and could be used to reveal unaccessible dynamics with standards OECTs.

## 2.3. Short-term memory effect engineering with dendritic structures.

Capacitive response in OECTs corresponds to the ability of OMIECs to accumulate ionic charges inside the bulk of the material and has been used to define the notion of volumetric capacitance.[25] Interestingly, volumetric capacitance and the associated ionic dynamics when gate voltage is charging / discharging the OMIECs have been used to mimic short-term memory effect in OECTs mimicking synaptic plasticity of biological synapses. Here, we were interested in using the non-ideal capacitive response to engineer short-term memory effects in dendritic OECTs. Square-shape pulses in between -0,4 and 0.4 V and 10 s duration were applied at the G terminal of the OECT while a constant SD bias of 100 mV was recording the current response (figure 3a). Gate current contribution was systematically removed from the SD current in order to analyze only the electronic response of the device. For low $f_p$, dendritic OECTs displayed non-symmetrical response for both positive and negative voltages (figure 3c) suggesting a preferential doping of the PEDOT material (i.e. either negative ions injection in or positive ions removal from the bulk of the dendrites). This was consistent with the lower conductivity suggesting lower doping level at low $f_p$. At the opposite, high $f_p$ dendritic OECTs (figure 3c) displayed equivalent response in magnitude at positive and negative voltages consistent with higher doping level from conductivity modeling. A second important feature evidenced in figure 3f was the non-monotonic response of charging for high gate voltage potential. This effect implied a competitive phenomenon other than the ganting effect taking place at $V_p < -0,3$ V, inducing a decrease of current for $t_p > 1s$. Possible reasons for such non-linear phenomenon with the voltage can lie at the molecular level (possibly Faradaic processes in the presence of benzoquinone and hydroquinone which was electrogenerated in parallel of the dendritic growth) or/and at the material level (electroactivity of PEDOT:PSS and ability to stretch or schrink with the voltage polarity). Interestingly, such non-monotonic response has been reported in short-term plasticity of biological synapses[26] and cannot be emulated with discrete electrostatic capacitor elements.. This result suggest that considering complex electro-chemical response of the whole system (i.e. dendritic OECT and electrolyte) could pave the way to new strategies to mimic complex biological synapses' dynamics.

Temporal footprints of both charging and discharging were subsequently analyzed (figure 3 e-g) in order to evaluate the short-term memory effect obtained with dendritic OECTs. As the morphologies have variable local widths represented with multiple cylinders, thus, variable local capacitance and in turn, combination of time scales can be expected from dendritic OECTs. Since time response cannot be described with a simple exponential, we report only the linear fitting of the charging / discharging curves at $t < 1s$. For both charging and discharging, increasing $f_p$ resulted in shorter time constant in agreement with the lower ionic capacity of thinner and less arborized dendrites. No clear tendency was observed for a given $f_p$ except for the lowest $f_p$ one which suggested a different dynamics for charging at positive and negative voltages. Overall, dendritic shape engineering with $f_p$ resulted in the possibility to tune the short-term response of dendritic OECTs and could be a useful strategy for synaptic plasticity engineering. For instance, the realization of short-term plasticity can be used to create transfer function in neuromorphic computation. Since the amplitude of the signal is dependent on the inter pulse duration, the OECTs can offer dynamical gain in the signal, wherein signal of various frequencies can be suppressed with different effectiveness. The variable values of time scales of the dendrites can also be explored for the applications of reservoir computing, where the multiple time scales of dendrites can be used to store the time history of the signal needed for efficient classification.[27]

## 2.4. Long-term memory effect

Learning in biological neural networks relies on a combination of both short-term and long-term memory effects.[28] Such long-term memory effects have been recently implemented in OECTs.[29] This proposition relies on the capacity of the gate electrode and channel of the OECT to permanently store charges when electronic circuit (i.e. S, D,and G electrical connections) is in an open circuit configuration. This mechanism is an interesting option for implementing long-term heterosynaptic plasticity that we implemented here in dendritic OECTs. Figure 4a shows how bipolar electropolymerization can be used to create additional connections in a dendritic network. Same bipolar signal as previously was applied to the gate terminal with S and D terminal grounded. Effective dendritic growth from the Au gate terminal was observed showing the possibility to engineer multiples nodes in a 3D configuration. No significant changes in the dendritic SD connections was observed, but additional electropolymerization on it cannot be rule out. Figure 4b and 4c show the programming strategy used to induce long-term memory effect into the SD dendritic connection. During programming, the gate potential was swept in between 0 and (+/-) $V_G$ in order to induce potentiation or depression of the channel. In between programming event, the gate terminal was electrically disconnected from the SD terminals through high impedance switching matrix in order to avoid electronic charges exchange. A constant SD potential of 0.1 V was maintained in order to read the dendritic OECT channel conductance. Figure 4d presents the evolution of the conductance during successive programming event with increased amplitude. While relaxation is still evident in between the positive and negative polarity programming, much longer retention was observed with respect to short-term dynamics reported in figure 3 (note that the timing in between two programming events was > 30 s). Relaxation could be explained by leakage in the electronic circuit but could be potentially circumvented by using larger ON/OFF ratio switches as in[30] or by engineering ionic intercalation in the dendritic materials such as in.[31] The long-term memory phenomena between the two dendrite structures, controlled by the third dendrite morphology, mimics heterosynaptic mechanisms, wherein, astrocytes controls the synaptic plasticity value between two neighboring neurons.[32]

## 3. Conclusion

We have demonstrated in this paper how bipolar electropolymerization could open the way to new neuromorphic engineering strategies. The bottom-up nature of electropolymerization combined with 3D nature of the dendrites was used to implement structural plasticity, which is reminiscent of neurogenesis in biological networks. We note that the integration aspect of these 3D objects still remains an unexplored direction that will require to rethink the standard circuit engineering approaches. Interestingly, 3D microfabric with printable structures could be an interesting technic that could be associated to bottom-up electropolymerization for creating new 3D circuit concepts. Furthermore, dendritic PEDOT:PSS fibers have shown unique electrical characteristics with capacitive response largely dominated by a CPE element. Combination of this unconventional circuit element with OECT properties of the dendritic fibers is offering new options for designing neuromorphic hardware, notably by offering unconventional temporal responses. In this work, such dynamics associated with electrolyte properties were used to demonstrate short-term plasticity. Future work should explore in more details how various electrolyte composition can impact the dendritic growth and the material properties of the organic fibers. Finally, the combination of structural plasticity with synaptic plasticity should be used for demonstrating actual neuromorphic functions and to evaluate the benefit of bottom-up engineering of neuromorphic circuits.

## 4. Experimental Section/Methods

**Materials and instrumentation.** Dendrites' growth was carried out by alternating current bipolar electropolymerization technique in an aqueous solution containing 1 mM of poly(sodium-4-styrene sulfonate) (NaPSS), 10 mM of 3,4-ethylenedioxythiophene (EDOT) and 10 mM of 1,4-benzoquinone (BQ). All chemicals were purchased from Sigma Aldrich and used without any further modification. Two 25 $\mu$m-diameter Au wires (purchased from GoodFellow, Cambridge, UK) serving as as working and grounded electrodes were immersed into a 20 $\mu$l electrolyte's drop placed onto a Parylene C covered glass substrate. Both electrodes were equally elevated on a controlled height from the substrate and positioned at a distance of 240 $\mu$m from each other. Bipolar square-wave signals of 10Vpp amplitude, fixed duty cycle and V offset values were generated from a 50MS/s Dual-Channel Arbitrary Waveform Generator (Tabor Electronics) with consistent variation of applied frequency (f) parameter. Each growth of individual dendrite morphology was carried out with unused gold wires and daily-prepared solutions. Electropolymerization process was recorded with a VGA CCD color Camera (HITACHI Kokusai Electric Inc).

**Electrical characterization** was conducted on Agilent B1500A Semiconductor Device Analyzer coupled with B2201A Switching Matrix.

**Electrochemical impedance spectroscopy** (EIS) was performed from 1 MHz to 10 mHz on a system from two gold wires serving as $V_{in}$ and Ag/AgCl gate serving as $V_{out}$ with using a Solartron Analytical (Ametek) impedance analyzer. After electropolymerization dendrites' functionalizations obtained at different frequencies were studied by impedance spectroscopy with a constant bias ($V_{DC} = 0$ V, $Va = 10$ mVrms) in the same electrolyte environment containing supporting electrolyte NaPSS as well as EDOT, BQ and HQ. Additionally, at the same conditions impedance spectroscopy was implemented for microfabricated OECT devices presented on a substrate by a pair of gold electrodes and a layer of ePEDOT in between.

**Circuit Impedance Modeling** was realized through fitting of the raw data spectra without digital-filter preprocessing. As an instrument there was employed an open-source EIS Spectrum Analyzer software.[**Bondarenko2005**]. The RC parameter fitting was manually adjusted at the visual appreciation of the simultaneous comparison of the Nyquist plots, Bode's modulus and Bode's phase plots.

**Growth of 3$^{rd}$ dendrite.** Preliminary in between 2 wires a structure of dendrites at 10Vpp, 80Hz, 50% duty cycle and 0 V offset was formed. After its connection, 3rd Au wire was immersed into the system, lifted up at the same height and placed at 240 μm distance from a place of dendrites completion. Similar aforementioned signal was applied from the 3rd Au wire electrode while one of the connected electrodes was grounded and another one left floating. For signal generation 50MS/s Dual-Channel Arbitrary Waveform Generator (Tabor Electronics) was employed.

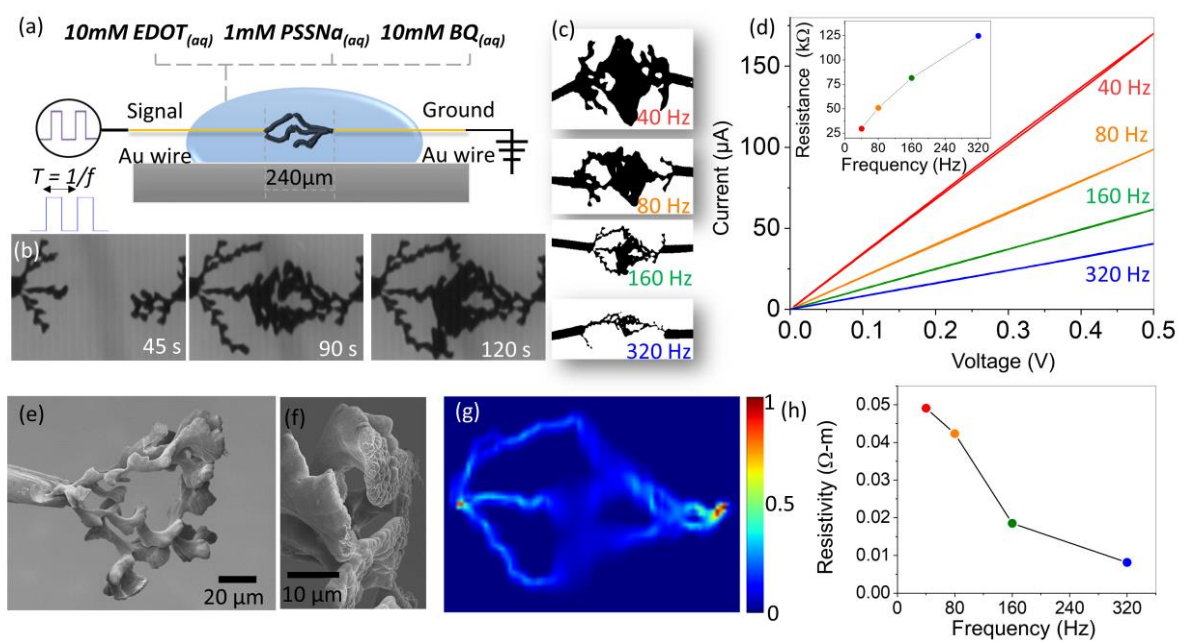

**Figure 1.** Structural plasticity with dendritic PEDOT:PSS fibers. (a) Schematic representation of the experimental set-up for dendritic growth with bipolar electro-polymerization. A periodic square signal of $V_{pp}$ with frequency $f_p$ of 40 Hz, 80 Hz, 160 Hz and 320 Hz is applied in between the two freestanding Au wires. (b) Temporal evolution of the formation of PEDOT dendrites with $f_p = 160$ Hz. (c) Comparison of the morphologies achieved for different $f_p$. (d) Current-Voltage characteristics of the different dendrites. (e,f) SEM images of dendrites grown at 80 Hz. (g) Normalized current density map for the dendrite obtained at $f_p = 160$ Hz based on image analysis and electrical simulations. (h) Calculated value of resistivity from the experimental resistance value and image's predicted resistance value.

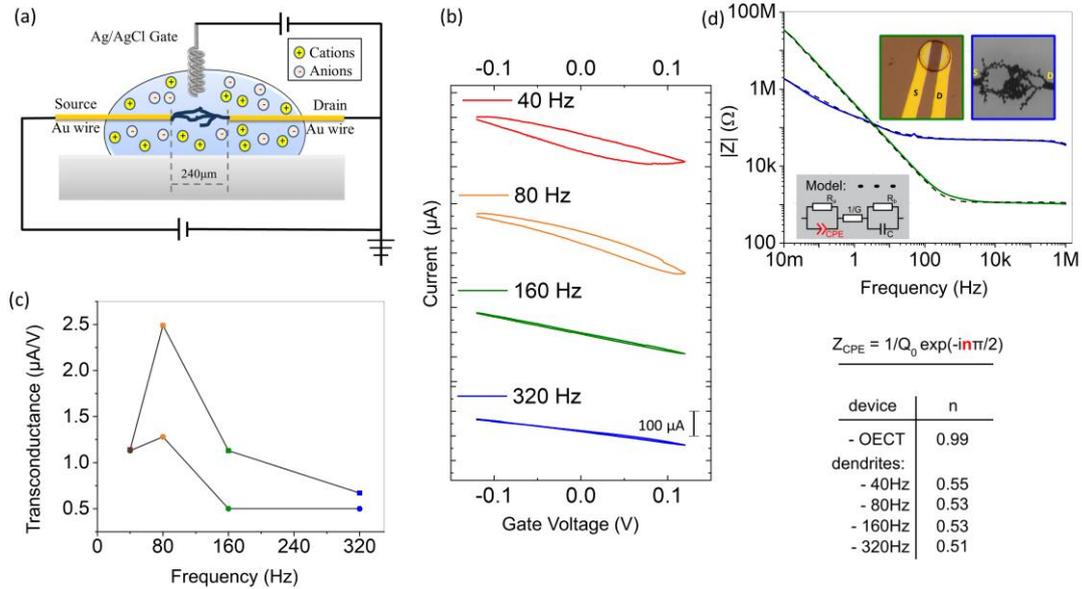

**Figure 2.** Electrical characteristics of dendritic OECTs. (a) Schematic of the OECT setup with Ag/AgCl as a gate electrode. (b) Transfer characteristic of the OECTs evidencing accumulation and depletion mode at negative and positive $V_G$, respectively. (c) Transconductance values fort the forward (square symbols) and backward (circle symbols) at $V_G = 0V$. (d) Impedance spectroscopy of standard OECT deposited by spin-coating and dendritic OECT at $f_p = 160$ Hz. The n value, which is 0 for a perfect resistor, 1 for ideal capacitor and 0.5 for Warburg element.

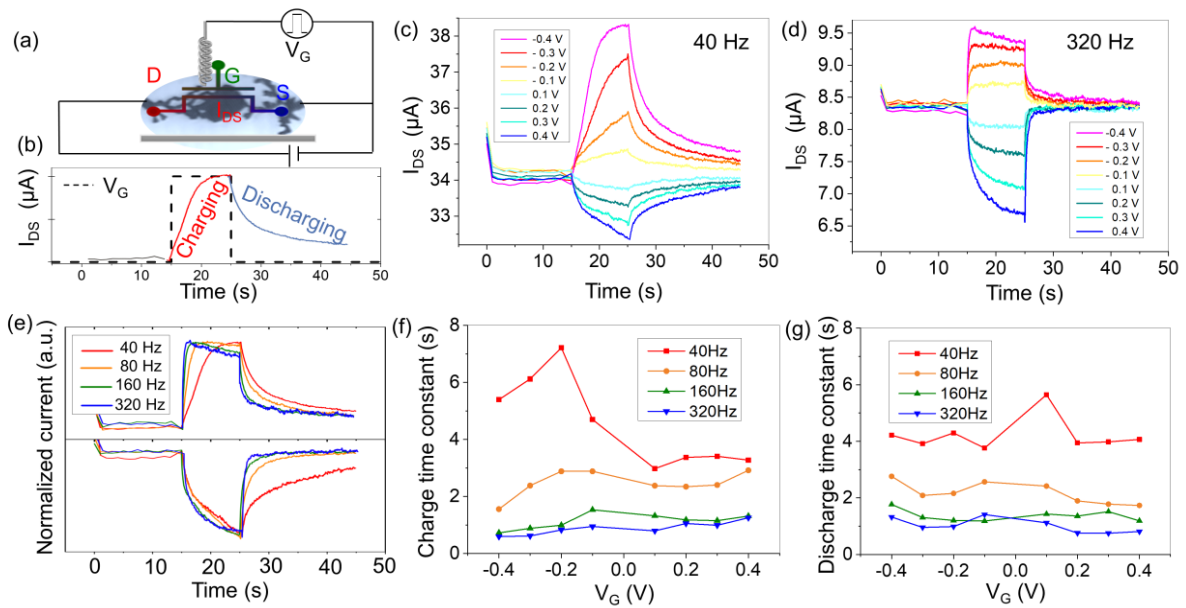

**Figure 3.** Short-term plasticity effect demonstrated for various dendritic morphologies. (a) Square-shape pusles of 10 s were applied to the gate with continuous recording of Source-Drain voltage of 0.1 V. (b) Typical SD current response to a square shape pulse. (c-d) Source-Drain current responses for dendritic OECTs grown at (c) $f_p = 40$Hz and (d) $f_p = 320$ Hz with pulse amplitude from -0.4 to 0.4 V with step of 0.1 V. Potentiation (depression) is observed at negative (positive) gate voltages. (e) Normalized responses for dendritic OECTs grown at 40 Hz, 80 Hz, 160 Hz, and 320 Hz with gate pulses of 0.4 and -0.4 V. (f-g) Variation in time constant of charging / discharging regions.

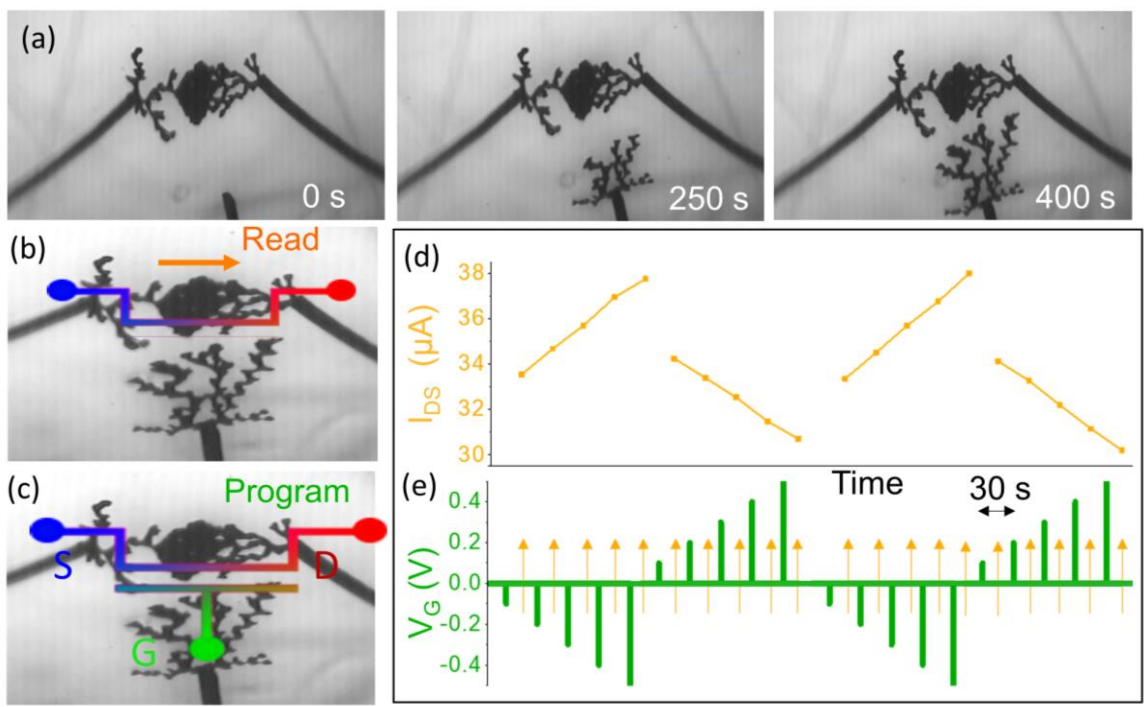

**Figure 4.** In-situ network evolution and long-term memory effects. (a) Formation of a third dendrite using the electropolymerization process with $f_p$ = 80Hz. (b) Schematic of the read operation with gate terminal disconnected. (c) Schematic of the programming of the dendritic OECTs with positive / negative sweep applied on the gate. (d) Successive program / read sequences with (e) $V_{sweep}$ increased from (+/-) 0.1 to 0.4 V (step of 0.1 V) in between each sequences. Long Term Potentiation (Depression) is obtained at negative (positive) bias. Time interval in between two successive programming was around 30 s.

**Acknowledgements**

K. Janzakova, M. Ghazal and A. Kumar contributed equally to this work.

We thank the RENATECH network and the engineers from IEMN for their support. F.A. thanks Yann Beilliard for their fruitful discussions.

This work is founded by ERC-CoG IONOS project #773228.

Received: ((will be filled in by the editorial staff))

Revised: ((will be filled in by the editorial staff))

Published online: ((will be filled in by the editorial staff))

**ToC**

We show in this paper how bipolar electropolymerization can be used to implement structural plasticity, a key ingredient from biological networks known as neurogenesis. Resulting dendritic PEDOT:PSS fibers are further used as organic electrochemical transistors to demonstrate short-term and long-term memory effects.

K. Janzakova, M. Ghazal, Dr. A. Kumar, Dr. Y. Coffinier, Dr. S. Pecqueur, Dr. F. Alibart

Dendritic organic electrochemical transistors grown by electropolymerization for 3D neuromorphic engineering

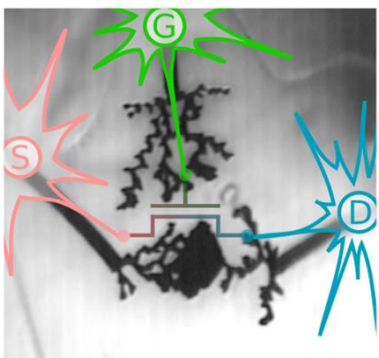

# Supporting Information

**Dendritic organic electrochemical transistors grown by electropolymerization for 3D neuromorphic Engineering**

*Kamila Janzakova, Mahdi Ghazal, Ankush Kumar, Yannick Coffinier, Sébastien Pecqueur\*, and Fabien Alibart\**

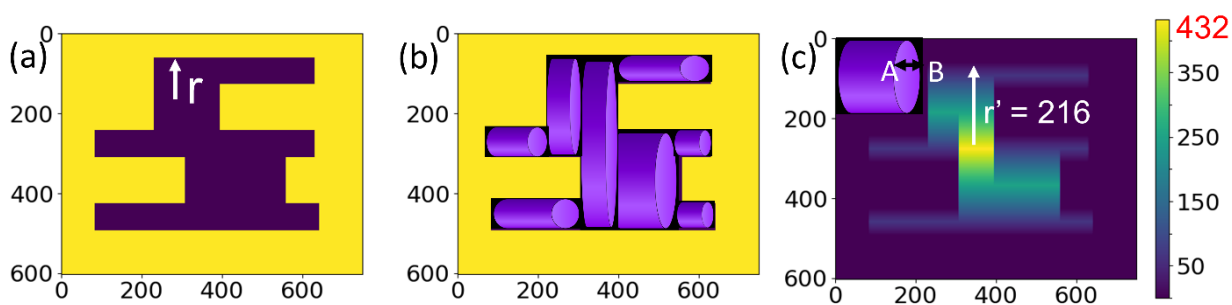

Figure S1: (a) An example of a 2d image, (b) reconstructed 3d image based on cylindrical shapes, (c) 3d profile of the image.

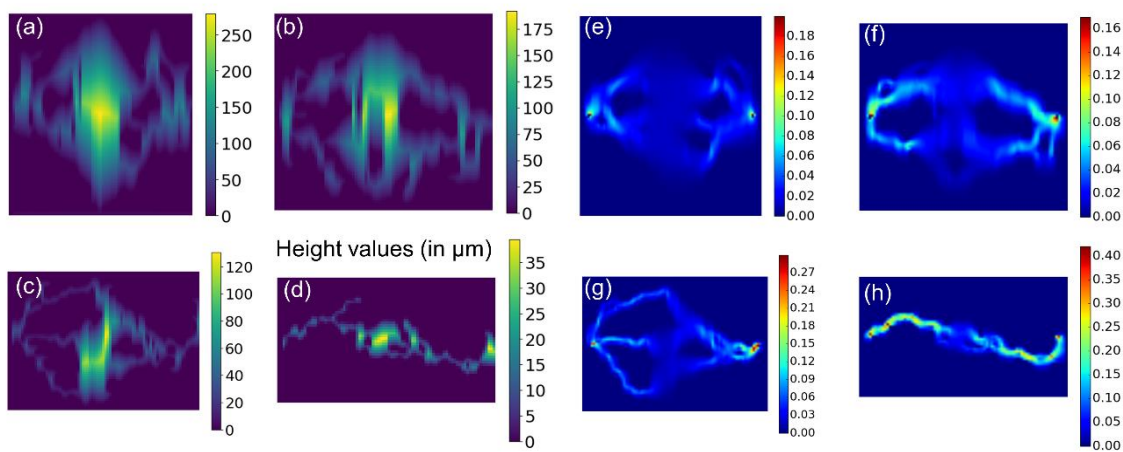

Figure S2: (a-d) 3-d morphology of dendrites based on multiple cylinder assumption and corresponding (e-h) normalized current density map for the dendrite obtained at fp = 40-320 Hz based on image analysis and electrical simulations.

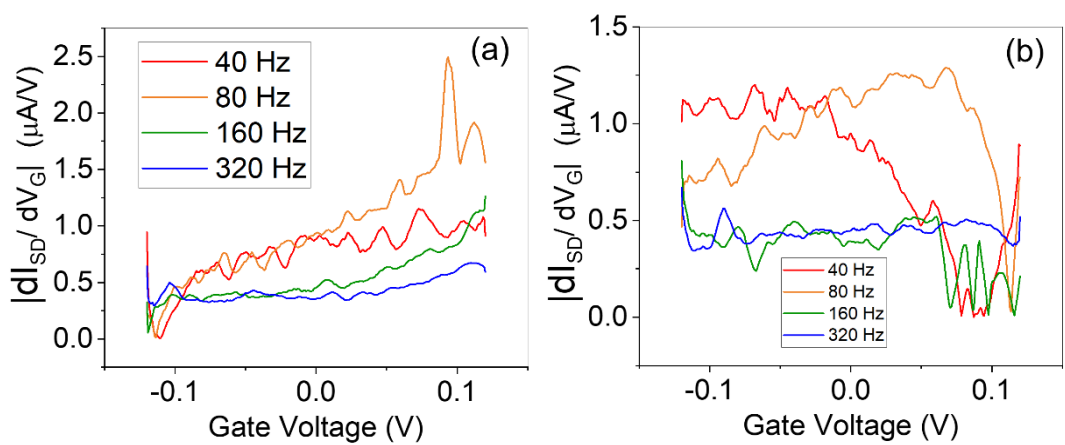

Figure S3: Variation in the derivative of source-drain current with the applied gate voltage during the (a) increasing and (b) decreasing values of the gate voltage.

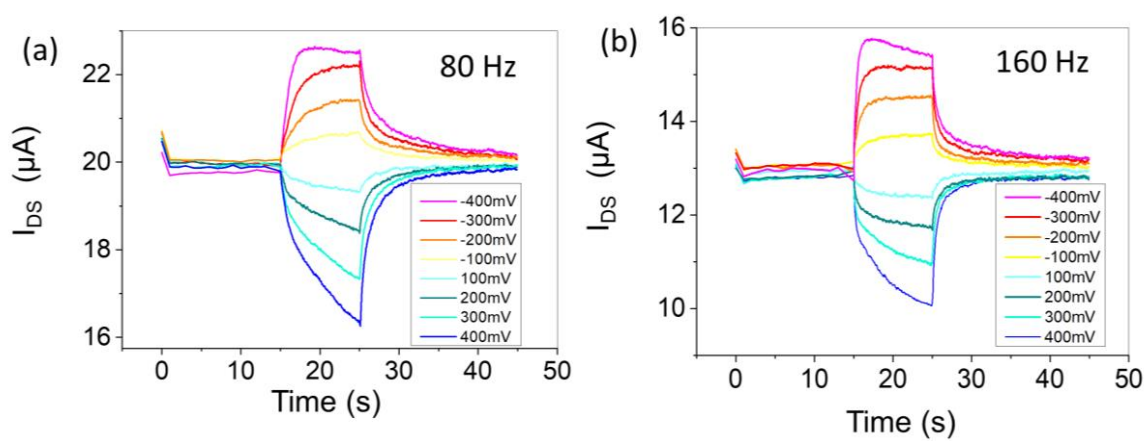

Figure S4: Source-Drain current responses for dendritic OECTs grown at (a) fp = 80Hz and (d) fp = 160 Hz with pulse amplitude from -0.4 to 0.4 V.